\documentstyle[12pt]{article}
\begin{document}
\hfill{PUTP-94-20}
\vspace{10mm}
\begin{center}
{\bf GLUONIC AND LEPTONIC DECAYS OF HEAVY QUARKONIA
 AND THE DETERMINATION OF $\alpha_s(m_c)$ AND $\alpha_s(m_b)$}
\vspace{15mm}\\

{Kuang-Ta CHAO$^{1,2}$~~~~ Han-Wen HUANG$^{2}$~~~~ Yu-Quan LIU$^{2}$}\\
\vspace{4mm}
1.{\it CCAST (World Laboratory),~Beijing 100080, P.R.China}\\
2.{\it Department of Physics, ~Peking University,
         ~Beijing 100871, P.R.China}
\end{center}
\vspace{8mm}
\begin{abstract}
QCD running coupling constant $\alpha_s(m_c)$ and
$\alpha_s(m_b)$ are determined from heavy quarkonia
$c\overline{c}$ and $b\overline{b}$
decays. The decay rates of $V\rightarrow 3g$ and $V\rightarrow e^+ e^-$
for $V=J/\psi$ and $\Upsilon$ are estimated by taking into account both
relativistic and QCD radiative corrections. The decay amplitudes
are derived in the Bethe-Salpeter formalism, and the decay rates are
estimated by using the meson wavefunctions which are obtained
with a QCD-inspired inter-quark potential.
For the $V\rightarrow 3g$ decay
we find the relativistic
correction to be very large and to severely suppress the decay rate.
Using the experimental values of
ratio $R_g\equiv \frac {\Gamma (V\longrightarrow 3g)}%
{\Gamma (V\longrightarrow e^{+}e^{-})}\approx 10,~32 $  for $V=J/\psi,
 ~\Upsilon$ respectively,
and the calculated widths , we find
$\alpha_{s}(m_c)=0.29\pm 0.02 $ and $\alpha_s(m_b)=0.20\pm 0.02$.
These values for the QCD running coupling
constant are substantially enhanced, as compared with the
ones obtained without relativistic corrections, and are consistent with the
QCD scale parameter $\Lambda_{\overline {MS}}^{(4)}%
\approx 200MeV$. We also find that these results are mainly due to
kinematic corrections and not sensitive to the dynamical models.
\end{abstract}
\vspace{8mm}

\eject
\section {\normalsize\bf INTRODUCTION}
\indent

Determining the QCD running coupling constant $\alpha_s$
at different energy scales is very important in the verification of
the fundamental theory of strong interaction. Among others the heavy
quarkonia decays may provide very useful information for $\alpha_s$
at the heavy quark mass scale. Decay rates of heavy quarkonia
in the nonrelativistic limit
with QCD radiative corrections have been studied (see, e.g., refs.[1,2,3]).
However, the decay rates of many
processes are subject to substantial relativictic corrections.
In particular, the rate of $J/\psi\rightarrow 3g$ and accordingly the
determination of $\alpha_{s}(m_c)$ depend rather
crucially on the relativistic corrections. Presently the ratios of
gluonic to leptonic widths of $J/\psi$ and $\Upsilon$ have been
precisely measured in experiment. However, if using the non-relativistic
expressions for decay widths which are proportional to $|R_S(0)|^2$,
where $|R_S(0)|$ is the radial wavefunction at the origin, and
comparing them with the corresponding experimental values,
one would get$^{[2,3]}$
$\alpha_s(m_c)=0.19$ and $\alpha_s(m_b)=0.17$. They are substantially
smaller than the expected values determined from other experimental
results or
the QCD scale parameter (for a review, see ref.[4]).
Theoretically, the difficulty is mainly due to the large relativistic
effects on the decay widths.
It is well known that there are at least two important
sources of relativistic effects for these
processes, one is from the kinematical corrections to decay amplitudes,
the other is from the bound state
wavefunction corrections which are concerned with the dynamical effect of
quark-antiquark interactions. In ref.[5],
the first order
relativistic corrections were considered based
on a phenomenological model. But only the relativistic
correction originated from kinematics was discussed
and no explicit methods with dynamical considerations were given
to calculate the decay widths. As a result of lack of reliable
estimates for relativistic corrections to the decay widths
in the determination of $\alpha_s(m_c)$ and $\alpha_s(m_b)$,
either an arbitrary choice for the $v^2/c^2$ term is made to get
these coupling constants enhanced$^{[3]}$,
or the $v^2/c^2$ term is arbitrarily neglected but a large
effective mass for the gluons is introduced$^{[6]}$.
Apparently, a better estimate
rather than arbitrary guesses for the relativistic
corrections is indeed needed, though this is certainly a very difficult
task without a deep understanding of quark confinement.

In this paper, as an attempt to tackle this problem, we will use
the Bethe-Salpeter (BS) formalism to derive the decay amplitudes
and to calculate the decay widths of $V\rightarrow
e^{+}e^{-}$ and $V\rightarrow 3g$, where $V$ is a vector heavy
quarkonium state. In this approach, the meson is considered as a bound
state consist of a pair of constitute quark and antiquark, and described
by the BS wavefunction which
satisfies the BS equation. A phenomenological QCD-inspired interquark
potential will be used to solve for the wavefunctions.
These may allow us to give the expressions which
take into account both relativistic and QCD radiative corrections
to next-to-leading order. We may then estimate both kinematical
and dynamical relativistic corrections to the decay widths, and
determine the QCD coupling constant from the calculated ratios of
gluonic to leptonic decay widths.
The remainder of this paper is organized
as follows. In Sec.2 we study the relativistic corrections
to $V\rightarrow e^{+}e^{-}$ and $V\rightarrow 3g$ widths. In Sec.3
we discuss the bound state wavefunctions and determine
the coupling constants $\alpha_s(m_c)$ and $\alpha_s(m_b)$
by using the experimental values of the gluonic
and leptonic widths of $J/\psi$ and $\Upsilon$. A summary and
discussion will be given in the last section.

\section {\normalsize\bf RELATIVISTIC CORRECTIONS TO $V\rightarrow e^+ e^-$
AND $V\rightarrow 3g$}
\indent

(1). The $V\rightarrow e^+ e^-$ decay

We first consider the
leptonic decay for $V=J/\psi$ and $\Upsilon$ (see also
ref.[7]).
This process proceeds
via the $Q\overline Q$ annihilation.
Define the Bethe-Salpeter wavefunction, in general,
for a $Q_1\bar Q_2$ bound state $|P>$
with overall mass $M$ and momentum $P=(\sqrt{{\vec P}^2+M^2},~{\vec P})$
\begin{equation}
\chi(x_1,x_2)=<0|T\psi_1(x_1)\bar\psi_2(x_2)|P>,
\end{equation}
where $T$ represents time-order product, and transform it into momentum
space
\begin{equation}
\chi_P(q)=e^{-iP\cdot X}\int d^4 x e^{-iq\cdot x}
\chi(x_1,x_2).
\end{equation}
Here $q_1~(m_1)$ is the quark momentum~(mass), $q_2~(m_2)$ the
antiquark momentum~(mass), and
$q$ the relative momentum,
\begin{eqnarray}
X=\eta_1 x_1+\eta_2 x_2, \ \ x=x_1-x_2,\\ \nonumber
P=q_1+q_2, \ \ q=\eta_2q_1-\eta_1q_2,
\end{eqnarray}
where $\eta_i={m_i\over{(m_1+m_2)}} ~(i=1,2)$.

In this formalism the quarkonium annilination matrix elements
can be written as
\begin{equation}
\langle 0\mid \overline{Q}I Q\mid P\rangle =\int d^4qTr\left[
I(q,P)\chi _P(q)\right],
\end{equation}
where $I(q,P)$ is the interaction vertex of the $Q\bar Q$ with other
fields (e.g., the photons or gluons ) which, in general, may also depend
on the variable $q^0$. If $I(q,P)$ is independent of
$q^0$, Eq.~(4) can be written as
\begin{equation}
\langle 0\mid \overline{Q}I Q\mid P\rangle
=\int d^3qTr[I
(\stackrel{\rightharpoonup }{q},P)\Phi_
{P}(\stackrel{\rightharpoonup }{q})],
\end{equation}
where
\begin{equation}
\Phi _{P}(\stackrel{\rightharpoonup }{q})
=\int dq^0 \chi_P(q)
\end{equation}
is the three dimensional BS wave function of the $Q\bar Q$ meson.
In the BS formalism the wavefunctions in the meson rest frame,
where ${\vec q_1}=-{\vec q_2}={\vec q},~P=(M,0)$, can be expressed as
\begin{eqnarray}
&&\Phi _{P}^{0^{-}}( \stackrel{%
\rightharpoonup }{q}) =\Lambda _{+}^1({\vec q})\gamma ^0( 1+\gamma ^0
) \gamma _5\gamma ^0\Lambda _{-
}^2(-{\vec q})\varphi(\stackrel{\rightharpoonup }{q}),\nonumber \\
&&\Phi _{P}^{1^{-}}( \stackrel{%
\rightharpoonup }{q}) =\Lambda _{+}^1({\vec q})\gamma ^0( 1+\gamma ^0
) \rlap/e\gamma ^0\Lambda _{-}^2(-{\vec q})
f(\stackrel{\rightharpoonup }{q}),
\end{eqnarray}
where $\Phi _{P}^{0^{-}}( \stackrel{%
\rightharpoonup }{q})$, and
$\Phi _{P}^{1^{-}}( \stackrel{%
\rightharpoonup }{q})$ represent the three dimensional wave
functions of $0^-$ and $1^-$ mesons respectively,
$\rlap/e=e_{\mu}\gamma^{\mu}$, $e_{\mu}$ is the polarization
vector of $1^-$ meson, $\varphi$ and $f$ are scalar functions which can
be obtained by solving the BS equation for $0^-$
and $1^-$ mesons, and
$\Lambda_{+} (\Lambda_{-})$ are the positive (negative) energy
projector operators
\begin{eqnarray}
&&\Lambda^{1}_{+}(\vec q)=\Lambda_{+}({\vec q}_1)=\frac 1{2E}(E+\gamma ^0%
{\vec \gamma}\cdot{\vec q}_1+m\gamma ^0),\nonumber\\
&&\Lambda^{2}_{-}(-\vec q)=\Lambda_{-}({\vec q}_2)=\frac 1{2E}(E-\gamma ^0%
{\vec \gamma}\cdot{\vec q}_2-m\gamma ^0),\nonumber\\
&&E=\sqrt{{\vec q}^2+m^2}.
\end{eqnarray}

For process $V\rightarrow e^+e^-$
with the electron (positron) momentum $p_1(p_2)$ and helicity
$r_1(r_2)$,
\begin{equation}
I({\vec q},P)=-ie\gamma_{\mu},
\end{equation}
and the amplitude can be written as
\begin{equation}
T= e^2 e_Q\langle 0\mid \overline{Q}
\gamma_{\mu} Q\mid V \rangle \overline{u}_{r_1}(p_1)
\gamma^{\mu} v_{r_2}(p_2)\frac 1{M^2},
\end{equation}
where $M$ is the meson mass and $e_Q$ is the electric charge
of the quark $Q$ $(Q=c, b)$. Define the decay constant $f_V$ by
\begin{equation}
f_VMe_\mu\equiv \langle 0\mid \overline{Q}\gamma _\mu Q\mid V \rangle
=\int d^3 q~Tr[\gamma _\mu
\Phi _{P}(%
\stackrel{\rightharpoonup }{q})],
\end{equation}
where $e_{\mu}$ is the polarization vector of $V$ meson.
Then with (7) we can easily find
\begin{equation}
f_V=\frac{2\sqrt{3}}M\int d^3 q~(\frac{m+E}E-\frac{%
\stackrel{\rightharpoonup }{q}^2}{3E^2})
f(\stackrel{\rightharpoonup }{q}),
\end{equation}
where $E=\sqrt{\stackrel{\rightharpoonup }{q}^2+m^2}$,
and $\sqrt{3}$ is the color factor.
Summing over the polarizations of the final state and
averaging over that of the initial state, it is easy
to get the decay width
\begin{equation}
\Gamma (V \rightarrow e^{+}e^{-})=
\frac 43\pi \alpha ^2e_Q^2f_V^2/M.
\end{equation}
Including further the QCD radiative correction and assuming that
the radiative correction and relativistic correction can be
factorized, we then get the
following expression for the
decay width with both relativistic and QCD radiative corrections
\begin{equation}
\Gamma (V \rightarrow e^{+}e^{-})=
\frac 43\pi \alpha ^2e_Q^2f_V^2/M (1-\frac{16\alpha_s(m_Q)}{3\pi}).
\end{equation}
Expanding $f_V$ in terms
of $\stackrel{\rightharpoonup }{q}^2
/m^2$, to the first order we get
\begin{equation}
f_V=-\frac{4\sqrt{3}}M\int d
\stackrel{\rightharpoonup }{q}f(\stackrel{%
\rightharpoonup }{q})(1-\frac 5{12}
\frac{\stackrel{\rightharpoonup }{q}^2}{%
m^2}).
\end{equation}

2. The $V\rightarrow 3g$ decay.

We next study the hadronic decay for $V=J/\psi$ and $\Upsilon$.
We consider a $J^{PC}=1^{--}$ $Q\bar Q$ bound state
decaying into three gluons.
The decay width is given by
\begin{equation}
\Gamma={1\over{2M}}\int d\phi Z ,
\end{equation}
where the integration $\int d\phi$ is over the final-state phase space.
$Z$ is defined by
\begin{equation}
Z=\sum |T|^2,
\end{equation}
where $\sum$ represents summing over the polarizations of the final
state and averaging over that of the initial state. The decay
matrix element $T$ is
\begin{equation}
T=-ig_s^3 Tr(T_a T_b T_c)\int d^4 q Tr\chi_{P}(q)
I(k_1,k_2,k_3;q_1,q_2),
\end{equation}
where
\begin{eqnarray}
I(k_1,k_2,k_3;q_1,q_2)=&& \frac{\rlap/{\epsilon}_3
(\rlap/k3-\rlap/{q_2}+m)\rlap/{\epsilon}_2
(\rlap/{q_1}-\rlap/{k_1}+m)\rlap/{\epsilon}_1}
{((k_3-q_2)^2-m^2)((q_1-k_1)^2-m^2)}\nonumber\\
&& + ~all~ permutations~ of~ 1,2,3.
\end{eqnarray}
Here $k_i$ and $\epsilon_i$ ($i$=1,2,3)
represent the momenta and polarizations of
 the three gluons;
$T_a$,$T_b$,$T_c$ are the color $SU(3)$ matrices,
and $a,b,c$ are the color indices of the three gluons;
$q_1$ and $q_2$ are the momenta of quark and antiquark,
and their time components satisfy
$q^0_1+q^0_2=M$.
As usual, we assume$^{[1,5]}$
\begin{equation}
q^0=q^0_1-q^0_2=0,\ \  q^0_1=q^0_2=M/2.
\end{equation}
Thus $I(k_1,k_2,k_3,q_1,q_2)$ is independent of $q^0$, and
$T$ becomes
\begin{equation}
T =-ig_s^3 Tr(T_a T_b T_c)\int d^3 q
Tr\{\Phi_{P}^{1^{-}}({\vec q})I(k_1,k_2,k_3;\vec q)\}.
\end{equation}
Substituting (7) and (19) into (21) and (17),
we get the expression for $Z$.
In the extremely non-relativistic limit the dependence on $\vec q$ of
$\Lambda_{+}(\vec q)$, $\Lambda_{-}(\vec q)$ and $I(k_1,k_2,k_3;\vec q)$
in $Z$ is neglected and we obtain the usual non-relativistic
formula for $V\rightarrow 3g$.
In order to go beyond the leading order result we expand the matrix
element $T$ in terms of ${\vec q}^2/{m^2}$. After performing the
trace of $\gamma$ matrix, for $m=M/2$
we get
to the first order of $v^2/c^2$
\begin{eqnarray}
Z=& &\frac{160 {g_s}^6}{81 M^4}\int d^3 qd^3 q^{\prime}
\{A_0(x_1,x_2,x_3)(1+\frac{{\vec q}^2+{\vec q}^{\prime 2}}{6 m^2})
\nonumber\\
& & +A_1(x_1,x_2,x_3)\frac{{\vec q}^2+{\vec q}^{\prime 2}}{3 m^2}\}
f({\vec q})f({\vec q^{\prime}}),
\end{eqnarray}
where $f(\vec q)$ is the scalar
wavefunction of vector meson which comes from (7), and where
\begin{eqnarray}
A_0(x_1,x_2,x_3)=&&{16\over{x_1^2}}+{16\over{x_2^2}}+{32\over{x_1 x_2}}
-{32\over{x_1^2 x_2}}-{32\over{x_1 x_2^2}}
+{16\over{x_1^2 x_2^2}}\nonumber\\
&&+~(two~ other~ permutations),\nonumber\\
A_1(x_1,x_2,x_3)=&&-{16\over{x_1^2}}-{16\over{x_2^2}}-{41\over{x_1 x_2}}
+{12\over{x_1^2 x_2}}+{12\over{x_1 x_2^2}}
-{40\over{x_1^2 x_2^2}}\nonumber\\
&&-{8\over{x_1^3}}-{8\over{x_2^3}}
-{8\over{x_1^3 x_2}}-{8\over{x_1 x_2^3}}
+{48\over{x_1^3 x_2^2}}+{48\over{x_1^2 x_2^3}}
-{32\over{x_1^3 x_2^3}}\nonumber\\
&&+~(two~ other~ permutations),\nonumber\\
x_i=& &{2\omega_i\over{M}},~~~~  i=1,2,3.
\end{eqnarray}
where $\omega_i$ represents the gluonic energy.
The final-state phase space is given by
\begin{equation}
d\phi=(2\pi)^4\delta^4 (P-{\sum\limits_{i=1}^{3}}k_i)
{\prod\limits_{i=1}^{3}}\frac{d^3 k_i}{(2\pi)^3 2\omega_i}~~.
\end{equation}
After performing the integration, we find the decay width of $V
\rightarrow 3g$ with relativistic corrections to be
\begin{equation}
\Gamma (V \rightarrow 3g)=
\frac {640({\pi}^{2}-9)\alpha_{s}^{3}(m_Q)}{81 M^3} g_{V}^2,
\end{equation}
 where to the first order relativistic correction $g_V$ is
 given by
\begin{eqnarray}
g_V&=&\int d^3 q~ [1-(\frac{(41{\pi}^2/3-48)}{32({\pi}^2
-9)}-1/6)\frac{{\vec q}^2}{m^2}]~f(\vec q)\nonumber\\
   &=&\int d^3 q~ [1-2.96\frac{{\vec q}^2}{m^2}]~f(\vec q).
\end{eqnarray}
This integral will diverge when the wavefunction $f$ is determined
by the inter-quark potential which is
Coulombic at short distances where one gluon-exchange dominates.
We could introduce a cut-off scale $\Lambda$ of order
$m(v/c)$ to regularize the integral. However, examining the
interaction vertex $I$ given in (19) we can
easily see that $g_V$ should be a convergent quantity if all higher
order corrections are taken into consideration, because
in (19) $I$ becomes inversely proportional to ${\vec q}^2$ as
${\vec q}^2\rightarrow\infty$. In the absence of calculations for
higher order corrections we will take the following expression for
the regulation of $g_V$
\begin{equation}
g_V\approx \int d^3 q~[1+2.96\frac{{\vec q}^2}{m^2}]^{-1}
 ~f(\vec q),
\end{equation}
which gives the same result as (26) to the first order relativistic
correction. Similarly, for the leptonic decay the first order expression
of decay constant
$f_V$ in (15) may also be regularized in the same manner, which differs
from the full expression (12) only by higher order terms, and the
difference turned out to be small.

In connection with (20), another scheme is to take the on-shell
condition, which assumes (see, e.g., ref.[5])
the quark and antiquark to be on the mass shell
\begin{equation}
q^0_1=q^0_2=M/2=E=\sqrt{m^2+{\vec q}^2}.
\end{equation}
The advantage of this assumption is that the gauge invariance
is maintained for the on-shell quarks
but at the price of treating the quark and
antiquark just as free particles in a bound state.
An apparent problem in this scheme is that if the quark mass takes
a fixed value (as in the case before) then ${\vec q}^2$
will be fixed but not weighted by the
wavefunction as in the usual bound state description.
In order to connect the decay process, which occurs at short
distances, where quarks are approximately on shell,
with the bound state wavefunction, which is mainly
determined by the long distance confinement force, we have to
make a compromise between the on-shell condition and the bound
state description.
We will expand the matrix element (21) and (17)
in terms of $\frac{{\vec q}^2}{E^2}=
\frac{{\vec q}^2}{(M/2)^2}$, and allow ${\vec q}^2$
(so the quark mass accordingly) to vary
in accordance with the bound state wavefunction $f(\vec q)$
which is to be determined by the long distance dynamics, or
phenomenologically by some dynamical models. With this
treatment, we get another expression for $Z$
\begin{eqnarray}
Z=& &\frac{160 {g_s}^6}{81 M^4}\int d^3 qd^3 q^{\prime}
\{A_0(x_1,x_2,x_3)(1-\frac{5{\vec q}^2+5{\vec q}^{\prime 2}}
{6 (M/2)^2})\nonumber\\
& &+A_1(x_1,x_2,x_3)\frac{{\vec q}^2+{\vec q}^{\prime 2}}{3 (M/2)^2}\}
f({\vec q})f({\vec q^{\prime}}),
\end{eqnarray}
where
\begin{eqnarray}
A_0(x_1,x_2,x_3)=& &{16\over{x_1^2}}+{16\over{x_2^2}}+{32\over{x_1 x_2}}
-{32\over{x_1^2 x_2}}-{32\over{x_1 x_2^2}}
+{16\over{x_1^2 x_2^2}}\nonumber\\
               & &+~(two ~other ~permutations),\nonumber\\
A_1(x_1,x_2,x_3)=& &{20\over{x_1^2}}+{20\over{x_2^2}}+{28\over{x_1 x_2}}
-{56\over{x_1^3 x_2}}-{56\over{x_1 x_2^3}}-{88\over{x_1^2 x_2^2}}
\nonumber\\
& &+{72\over{x_1^3 x_2^2}}+{72\over{x_1^2 x_2^3}}+{16\over{x_1^3}}
+{16\over{x_2^3}}-{32\over{x_1^3 x_2^3}}\nonumber\\
& &+ ~(two ~other ~permutations).
\end{eqnarray}
Then to the first order relativistic correction, we get
\begin{eqnarray}
g_V &=& \int d^3 q~f(\stackrel{%
\rightharpoonup }{q})[1-(\frac {36-\frac {7}{3}{\pi}^2}{8(\pi^2-
9)}+\frac {5}{6})\frac {\stackrel{\rightharpoonup}
{q}^2}{(M/2)^2}]  \nonumber\\
    & \approx & \int d^3 q~f(\stackrel{%
\rightharpoonup }{q})[1+2.7\frac {\stackrel{\rightharpoonup }%
{q}^2}{(M/2)^2}]^{-1},
\end{eqnarray}
where $g_V$ has been regulated in the same manner as (27).

Including futher the QCD radiative correction and assuming again
that the radiative and relativistic corrections can be factorized,
we then get the following
expression for the decay width with both relativistic and QCD
radiative corrections
\begin{equation}
\Gamma (V \rightarrow 3g)=
\frac {640({\pi}^{2}-9)\alpha_{s}^{3}(m_Q)}{81 M^3}
(1-C\frac{\alpha_s(m_Q)}{\pi})g_{V}^2,
\end{equation}
where $C=3.7(4.9)$ for $Q=c(b)^{[2,3]}$. In the extremely non-relativistic
limit, the ${\vec q}^2/{m^2}$ term is zero, and
the scalar wavefunction $f(q)$ satisfies the following relation
\begin{equation}
\int d^3 q~ f(\vec q)=\sqrt{M\over 4}~\psi(0),
\end{equation}
where $\psi(0)$ is the nonrelativistic
wavefunction at the origin in coordinate space, we then get the
well-known zero-order result$^{[2,3,5]}$
\begin{equation}
\Gamma(V\rightarrow 3g)=\frac{160({\pi}^2-9)\alpha^3_s(m_Q)}
{81 M^3}|\psi(0)|^2.
\end{equation}

With the two expressions (26) and (31), obtained in the two
different treatments concerning the on-shell condition, we find that
they both give very close results.
This may largely reduce the uncertainties in our calculations
associated with the
on-shell or off-shell descriptions for the quarks which
decay at short distances (therefore approximately
on shell) and
are bound together at large distances (therefore off shell).

Comparing (26) and (31) with (15), we see that the suppression due to
relativistic correction for $V\rightarrow 3g$ is much more
severe than for $V\rightarrow e^{+}e^{-}$. This result then
rules out the conjecture that the relativistic correction to
$J/\psi\rightarrow 3g$ may be neglibibly small$^{[6]}$.

\section {\normalsize\bf DETEMINATION OF $\alpha_s(m_Q)$ AND
BOUND STATE WAVE FUNCTIONS}
\indent

To calculate the widths of leptonic and gluonic decays,
we have to know the wavefunctions $f(\vec q)$ for $c\bar c$ and
$b\bar b$ states, which are determined mainly by the long
distance inter-quark dynamics. In the absence of a deep understanding
for quark confinement at present, we will follow a phenomenological
approach by using QCD inspired inter-quark potentials,
which are supported
by both lattice QCD calculations and heavy quark phenomenology,
as the interaction kernel in the BS equation. We begin with the
bound state BS equation$^{[8]}$ in momentum space
\begin{equation}
(\rlap/{q}_1-m_1)\chi_P(q)(\rlap/{q}_2+m_2)
={i\over{2\pi}}\int d^4 k G(P,q-k)\chi_P(k),
\end{equation}
where $q_1$ and $q_2$ represent the momenta of quark and
antiquark respectively,
$G(P,q-k)$ is the interaction kernel which dominates the inter-quark dynamics.
In solving Eq.(35), in order to avoid the notorious problem due to the
excitation of relative time variable we have to employ the "instantaneous
approximation". Meanwhile, we will neglect negative energy projectors in the
quark propagators. We then get the reduced Salpeter equation$^{[8]}$
for the three dimensional BS wavefunction $\Phi_{P}({\vec q})$ defined
in (6)
\begin{equation}
\Phi_{P}({\vec q})={1\over{P^0-E_1-E_2}}\Lambda_{+}^1
\gamma^0\int d^3 k G( P,{\vec q}-{\vec k})
\Phi_{P}({\vec k})\gamma^0\Lambda^2_{-},
\end{equation}
where $G(P,{\vec q}-{\vec k})$ represents the
instantaneous potential. We employ the following
interquark potrntials including a long-ranged confinement potential
(Lorentz scalar) and a short-ranged one-gluon exchange potential
(Lorentz vector)$^{[7]}$
\begin{eqnarray}
&&{V(r)}={V_S(r)+\gamma_{\mu}\otimes\gamma^{\mu} V_V(r)},\nonumber \\
&&{V_S(r)}={\lambda r\frac {(1-e^{-\alpha r})}{\alpha
r}},\nonumber \\
&&{V_V(r)}=-{\frac 43}{\frac {\alpha_{s}(r)} r}e^{-\alpha r},
\end{eqnarray}
where the introduction of the factor $e^{-\alpha r}$ is to regulate
the infrared divergence and also to incorporate
the color screening
effects of the dynamical light quark pairs on the $Q\bar Q$
linear confinement potential$^{[9]}$.
In momentum space the potentials become$^{[7]}$
\begin{eqnarray}
&&G( \stackrel{\rightharpoonup }{p})=G_S( \stackrel{%
\rightharpoonup }{p}) +\gamma_{\mu}\otimes \gamma^{\mu}
G_V( \stackrel{\rightharpoonup
}{p}),\nonumber \\
&&G_S( \stackrel{\rightharpoonup }{p})=-\frac \lambda \alpha
\delta ^3( \stackrel{\rightharpoonup }{p})+\frac \lambda {\pi
^2}\frac 1{( \stackrel{\rightharpoonup }{p}^2+\alpha ^2)
^2},\nonumber \\
&&G_V( \stackrel{\rightharpoonup }{p})=-\frac 2{3\pi^2}
\frac {\alpha_{s}(\stackrel{\rightharpoonup
}{p})}{\stackrel{\rightharpoonup }{p}^2+\alpha ^2},
\end{eqnarray}
where $\alpha_{s}(\stackrel{\rightharpoonup }{p})$ is the
quark-gluon running
coupling constant and is assumed to become a constant of $O(1)$ as
${\stackrel{\rightharpoonup }{p}}^2\rightarrow 0$
\begin{equation}
\alpha _s( \stackrel{\rightharpoonup }{p}) =\frac{12\pi }{27}%
\frac 1{\ln ( a+{\stackrel{\rightharpoonup }{p}^2}/{\Lambda
_{QCD}^2%
}) }.
\end{equation}
The constants $\lambda$, $\alpha$, $a$ and $\Lambda_{QCD}$ are
the parameters
that characterize the potential. Substituting (7) and (38)
into Eq.(36), one derives the equation for the $1^-$ meson
wavefunction $f(\vec q)$ in the meson rest frame$^{[7]}$
\begin{eqnarray}
M f_1({\vec q}) &=& (E_1+E_2)f_1({\vec q}) \nonumber\\
& & -{1\over{4 E_1 E_2}}\int d^3 k (G_S({\vec q}-{\vec k})
-2G_V({\vec q}-{\vec k}))(E_1 m_2+E_2 m_1)f_1({\vec k}) \nonumber\\
& & -{{E_1+E_2}\over{4 E_1 E_2}}\int d^3 k G_S({\vec q}-{\vec k})
\frac{E_1 m_2+E_2 m_1}{m_1+m_2}f_1({\vec k}) \nonumber\\
& & +\frac{E_1 E_2-m_1 m_2+{\vec q}^2}{4E_1 E_2{\vec q}^2}
\int d^3 k(G_S({\vec q}-{\vec k})+4G_V({\vec q}-{\vec k}))
({\vec q}\cdot{\vec k})f_1({\vec k}) \nonumber\\
& & -\frac{E_1 m_2-E_2 m_1}{4E_1 E_2{\vec q}^2}\int d^3 k
(G_S({\vec q}-{\vec k})-2G_V({\vec q}-{\vec k}))({\vec q}
\cdot{\vec k})\frac{E_1-E_2}{m_1+m_2}f_1({\vec k}) \nonumber\\
& & -\frac{E_1+E_2-m_1-m_2}{2E_1 E_2{\vec q}^2}\int d^3 k
G_S({\vec q}-{\vec k})({\vec q}\cdot{\vec k})^2
{1\over{E_1+E_2+m_1+m_2}}f_1({\vec k}) \nonumber\\
& & -\frac{m_1+m_2}{E_1 E_2{\vec q}^2}\int d^3 k
G_V({\vec q}-{\vec k})({\vec q}\cdot{\vec k})^2
{1\over{E_1+E_2+m_1+m_2}}f_1({\vec k}),
\end{eqnarray}
where
\begin{equation}
f_1({\vec q})=-\frac{m_1+m_2+E_1+E_2}{4E_1 E_2}f({\vec q}).
\end{equation}
The normalization condition of the wavefunction $f({\vec q})$ reads
\begin{equation}
\int d^3 q \frac{2E_1 E_2(m_1+E_1)(m_2+E_2)}{(E_1+E_2+m_1+m_2)^2}
f_1^2({\vec q})={M\over{4\pi}^3}
\end{equation}
Eq.(40) looks very complicated since higher order terms are all involved.
To the leading order in the nonrelativistic limit, Eq.(40) is just the
ordinary nonrelativistic Schrodinger equation. To the first order of
$v^2/c^2$, Eq.(40) becomes the well known Breit equation with both
vector (one-gluon) exchange and scalar (confinement) exchange.

For the heavy quarkonium $c\bar c$ and $b\bar b$ systems,
$m_1=m_2=m$, Eq.(40) can be greatly simplified.
 By solving Eq.(40)
 we can find the wave functions for the $1^-$ mesons (see, e.g.,
 ref.[7]).

Substituting the obtained BS wave functions into (12) and (14),
we then get
\begin{equation}
\Gamma (J/\psi\longrightarrow e^+e^-)=5.6 keV,
\end{equation}
where we have used
\begin{eqnarray}
&&m_c=1.5GeV,\ \ \lambda=0.23GeV^2,\ \ \Lambda_{QCD}=0.18GeV,\nonumber\\
&&\alpha=0.06GeV,\ \ a=e=2.7183.
\end{eqnarray}
With these parameters the
$2S-1S$ spacing and $J/\psi-\eta_c$ splitting are required to fit the
data.
Our result is in agreement with the experimental value of
$\Gamma (J/\psi\longrightarrow e^+e^-)=5.36\pm 0.29 keV$
$^{[4]}$.
Here in above calculations the value of $\alpha_s (m_c)$ in the
QCD radiative
correction factor in (14) is chosen to be 0.29$^{[3]}$, which is also
consistent with our determination from the ratio of $BR(J/\psi
\rightarrow 3g)$ to $BR(J/\psi\rightarrow e^+e^-)$ (see below).

Using
the experimental data$^{[4]}$
\begin{equation}
R_g\equiv \frac {\Gamma (J/\psi\rightarrow 3g)}%
{\Gamma (J/\psi\rightarrow e^{+}e^{-})}\approx 10,
\end{equation}
and the calculated widths from (14) and (31), (32), we find
\begin{equation}
\alpha_{s}(m_c)=0.29\pm 0.02,
\end{equation}
as campared with the value without relativistic corrections
(but with QCD radiative corrections)
\begin{equation}
\alpha_{s}^{0}(m_c)=0.19\pm 0.02.
\end{equation}
Clearly, it is the strong suppression due to the relativistic
correction to $J/\psi\rightarrow 3g$ that enhances the value of
$\alpha_{s}(m_c)$ and then makes it consistent with the
determined QCD scale parameter $\Lambda_{\overline {MS}}^{(4)}
\approx 200MeV$ ( for a review see ref.[4] ). ( Note that a slightly
larger value for $\alpha_s(m_c)$ could be obtained if using (27)
rather than (31). )

In comparision we have chosen two other groups of parameters and
solved the BS equation for $c\bar c$ states by requiring again
both their  $2S-1S$ spacing
and $J/\psi -\eta_c$ splitting  fitting the data. By the same
procedure we get two values of $\alpha_s(m_c)$ corresponding
to the obtained two new wavefunctions of $J/\psi$. With
$m_c=1.4GeV, \lambda=0.24GeV^2$ and other parameters unchanged
 (the heavy quarkonia mass spectra are not sensitive to $a$ and
 $\alpha$ for $\alpha\leq 0.06 GeV)$ we get
$\alpha_s(m_c)=0.29.$
With
$m_c=1.6GeV, \lambda=0.22GeV^2$
and other parameters unchanged we get
$\alpha_s(m_c)=0.28.$

Meanwhile, we have also solved the nonrelativistic Schr$\ddot{o}$dinger
equation for the scalar function $f$ in the nonrelativistic
limit by using the same potentials and parametres as (38) and (44),
and performing the same calculations, we obtain
\begin{equation}
\alpha_s(m_c)=0.27.
\end{equation}
Note that in this limit the spin symmetry between the $1^-$ and
$0^-$ mesons is restored and the relativistic correction from the
dynamical sourse is eliminated.

In order to see further the sensitivity of the value of $\alpha_s(m_c)$
to the wavefunctions, we have also tried the Gaussian function
\begin{equation}
f(q)=N~exp~(-\frac{4{\vec q}^2}{3q^{2}_0}),
\end{equation}
where $N$ is the normalzation factor, and $q^{2}_{0}$ is the mean
value of the momentum squared of the quark inside the meson,
which may be roughly estimated by using the scaling
law $q^{2}_0=mC/2 ~(C=0.73 GeV)$ found for heavy quarkonia
(see, e.g., ref.[10]). Then with $m_c=1.5GeV$ we find that
while the gluonic and leptonic decay widths both become smaller,
their ratio does not change very
much and gives $\alpha_s(m_c)=0.26$.

These results indicate that the determination of $\alpha_s(m_c)$
is not sensitive to the dynamical corrections to
the wavefunctions, and even not very sensitive to the form of
wavefunctions, and the most important contribution
comes from the kinematic corrections.

For the $b\overline b$ system, with a similar calculation for the
decay rates of
$\Upsilon\rightarrow 3g$ and $\Upsilon\rightarrow e^+e^-$ with both
relativistic
and QCD radiative corrections taken into account, and using the
observed value of the ratio$^{[4]}$
\begin{equation}
R_g\equiv \frac {\Gamma (\Upsilon\rightarrow 3g)}%
{\Gamma (\Upsilon\rightarrow e^{+}e^{-})}\approx 32,
\end{equation}
we find
\begin{equation}
\alpha_{s}(m_b)=0.20\pm 0.02,
\end {equation}
as compared with the value without relativistic  corrections
(but with QCD radiative corrections)
\begin{equation}
\alpha_{s}^{0}(m_b)=0.17\pm 0.02.
\end{equation}
The value given in (51) agrees with the expected value from the
QCD scale parameter.
Moreover, with both relativistic and QCD radiative corrections,
and $\alpha_s(m_b)=0.20$, we get$^{[7]}$
\begin{equation}
\Gamma(\Upsilon\rightarrow e^+e^-)=(1.3\pm0.2) KeV,
\end{equation}
which is also in agreement with data$^{[4]}$.

We may then conclude that by estimating the relativistic
corrections to the gluonic and leptonic decays of heavy quarkonia,
we find that the relativistic effects substantially
suppress the $V\rightarrow 3g$ decays, and consequently
the determined values of the QCD running coupling
constant at the heavy quark mass scale can get enhanced, and can be
consistent with
other theoretical and experimental expectations.

\section{\normalsize\bf  SUMMARY AND DISCUSSION}
\indent

In this paper we calculated the first order relativistic corrections
to $V\rightarrow e^{+}e^{-}$ and $V\rightarrow 3g$ for the vector heavy
quarkonia $V=J/\psi$ and $\Upsilon$, based on the BS formalism for the
decay amplitudes and bound state wavefunctions.
We derived the coefficients of the $v^2/c^2$ term in the decay rates
with two different treatments for the on-shell condition of bound quarks,
and obtained very similar values for the coefficients.
These results may largely reduce
the uncertaity in our calculations concerning the on-shell condition in
these decay processes. To maintain gauge invariance in the decay
amplitudes, the on-shell quarks are certainly better than off-shell quarks.
But within the quark-antiquark sector of the Fock states for a
heavy quarkonium, gauge invariance
may not be guaranteed and other higher Fock states may be needed.
Nevertheless, the closeness of the two large negatrive values
of the coefficients of the ${\vec q}^2/m^2$ term (see (26) and (31))
in the $V\rightarrow 3g$ widths within the two treatments may
encourage us to conclude that
 the relativistic effects suppress
$V\rightarrow 3g$ decays much more severely than
$V\rightarrow e^{+}e^{-}$ decays, therefore  can make the coupling
constants $\alpha_s(m_c)$ and $\alpha_s(m_b)$ substantially
enhanced, as compared
with the values obtained without relativistic corrections.

For a more accurate estimate of relativistic corrections, higher orders,
e.g., the $(v^2/c^2)^2$ terms should be taken into account, but this
is difficult in our approach, since to the higher orders many other
effects e.g. the quark pairs and the dynamical gluons may be involved.
In connection with higher order effects,
we have regularized the singularity associated with the derivative
of the wavefunctions at origin in a simple manner, which
is valid to the first order
of $v^2/c^2$  but is uncertain to higher order corrections.

We have solved the BS equation for the bound state wavefunctions
with QCD inspired inter-quark potentials (linear confinement
potential plus one gluon exchange potential)
as the BS kernel. With some popular parameters for the potentials
we obtained the wavefunctions and used them to calculate the
gluonic and leptonic decay widths and their ratios.
By comparing the BS wavefunctions with Schrodinger wavefunctions
and Gaussian-type wavefunctions,
we found that the ratios are not sensitive to the dynamical
relativistic effects on the wavefunctions, and even not very
sensitive to the form of wavefunctions.
We may then conclude that the ratios are insensitive to the
dynamical models,
and the relativistic effects on the ratios are mainly originated from
the kinematic parts of decay amplitudes.

As for the numerical result,
using the experimental values of
ratio $R_g\equiv \frac {\Gamma (V\longrightarrow 3g)}%
{\Gamma (V\longrightarrow e^{+}e^{-})}\approx 10,~32 $  for $V=J/\psi,
 ~\Upsilon$ respectively,
and the calculated widths, we found
$\alpha_{s}(m_c)=0.29\pm 0.02 $ and $\alpha_s(m_b)=0.20\pm 0.02$.
These values for the QCD running coupling
constant are consistent with the
QCD scale parameter $\Lambda_{\overline {MS}}^{(4)}%
\approx 200MeV$.

Recently, there have been some significant progresses in the study
of heavy quarkonium decays based on a more fundamental approach
of NRQCD (nonrelativistic QCD) (see refs.[11,12]). Many important
issues were clarified by this study. It will be interesting to
compare their results with ours in connection with
the gluonic and leptonic decay widths
and the determination of the strong coupling constant at the heavy
quark mass scales from these decays.

\vspace{10mm}

One of us (K.T.C.) would like to thank the hospitality of G. M. Prosperi
and N. Brambilla, the organizers of the International Conference on Quark
Confinement and the Hadron Spectrum, Como, Italy, 20-24 June 1994,
where this work was briefly reported. He also thanks E. Braaten for sending
him the Northwestern University preprints (ref.[11]).
This work was supported in part by the National Natural
Science Foundation of China, and the State Education
Commission of China.

\end{document}